\documentclass[12pt]{iopart}
\usepackage{graphicx} 
\usepackage{setstack}
\usepackage{iopams}

\newcommand{\dfrac}[2]{\frac{\displaystyle #1}{\displaystyle #2}}
\newcommand{\email}[1]{\ead{#1}}
\newcommand{\affiliation}[1]{\address{#1}}
\newcommand{\acknowledgments}{\ack}

\newcommand{\ie}{i.e.~}

\newcommand{\sss}[1]{{\scriptscriptstyle{#1}}}

\newcommand{\zero}{{\sss{0}}}

\newcommand{\calH}{\mathcal{H}}

\newcommand{\calS}{\mathcal{S}}
\newcommand{\calM}{\mathcal{M}}
\newcommand{\calW}{\mathcal{W}}
\newcommand{\calQ}{\mathcal{Q}}

\newcommand{\urad}{\mathrm{rad}}
\newcommand{\umat}{\mathrm{mat}}
\newcommand{\uc}{\mathrm{c}}
\newcommand{\ur}{\mathrm{r}}
\newcommand{\us}{\mathrm{s}}
\newcommand{\ud}{\mathrm{d}}

\newcommand{\uh}{\mathrm{h}}
\newcommand{\uphys}{\mathrm{phys}}

\newcommand{\aexp}{a}
\newcommand{\tphys}{t}

\newcommand{\corrini}{\ell_\uc}
\newcommand{\resini}{\ell_\ur}
\newcommand{\horizon}{d_\uh}
\newcommand{\horizonini}{d_{\uh_\zero}}
\newcommand{\enerlong}{\rho_{\infty}}
\newcommand{\enerloop}{\rho_{\circ}}
\newcommand{\enertot}{\rho}
\newcommand{\prestot}{P}
\newcommand{\masstot}{\calM}
\newcommand{\hubbleconf}{\calH}
\newcommand{\worktot}{\calW}
\newcommand{\losttot}{\calQ}

\newcommand{\frachor}{\alpha}

\newcommand{\frachorscal}{\frachor_\us}
\newcommand{\frachorlong}{\frachor_\infty}
\newcommand{\lphys}{l_\uphys}  
\newcommand{\tension}{U}      
\newcommand{\scaling}{\calS}
\newcommand{\numloop}{n}
\newcommand{\noconst}{C}
\newcommand{\const}{\noconst_\circ}
\newcommand{\power}{p}

\begin{document}

\title{Cosmological evolution of cosmic string loops}

\author{Christophe Ringeval}
\email{ringeval@fyma.ucl.ac.be}
\affiliation{Blackett Laboratory, Imperial College, Prince
Consort Road, London SW7 2AZ, United Kingdom\\ Theoretical and
Mathematical Physics Group, Centre for Particle Physics and
Phenomenology, Louvain University, 2 Chemin du Cyclotron, 1348
Louvain-la-Neuve, Belgium}

\author{Mairi Sakellariadou}
\email{mairi.sakellariadou@kcl.ac.uk}
\affiliation{Department of Physics, King's College, University of
London, Strand, London WC2R 2LS, United Kingdom}

\author{Fran\c{c}ois R.~Bouchet}
\email{bouchet@iap.fr}
\affiliation{Institut d'Astrophysique de Paris, 98bis boulevard Arago,
  75014 Paris, France}

\date{\today}

\begin{abstract}

The existence of a scaling evolution for cosmic string loops in an
expanding universe is demonstrated for the first time by means of
numerical simulations. In contrast with what is usually assumed, this
result does not rely on any gravitational back reaction effect and has
been observed for loops as small as a few thousandths the size of the
horizon. We give the energy and number densities of expected cosmic
string loops in both the radiation and matter eras.  Moreover, we
quantify previous claims on the influence of the network initial
conditions and the formation of numerically unresolved loops by
showing that they only concern a transient relaxation regime. Some
cosmological consequences are discussed.

\end{abstract}




Cosmic strings are line-like topological defects which may have formed
during phase transitions in the early universe. These defects are
primordial vacuum remnants associated with the spontaneous breakdown
of particle physics symmetries occurring during the cooling of the
universe~\cite{kibble:1976,kibble:1980}. If cosmic strings are formed
at the end of inflation~\cite{Jeannerot:2003qv}, they should still be
present in the observable universe and many studies have been devoted
to the understanding of their cosmological implications (see
Refs.~\cite{Vilenkin:2000,Hindmarsh:1994re} for a review and
references therein and thereto). Recently, cosmic strings have been of
particular interest since it was pointed out that Fundamental and
Dirichlet strings formed at the end of brane inflation could also play
a cosmic string-like role in
cosmology~\cite{Sarangi:2002yt,Copeland:2003bj,Dvali:2003zj}.

It is now part of the common lore that the existence of a string
network is cosmologically acceptable owing to the so-called ``scaling
regime'' of ``long strings''. Intersections between super-horizon
sized strings (the long, or infinite, strings) produce sub-horizon
sized loops, ensuring that the total energy density of long strings
scales with the cosmic time $t$ as $1/\tphys^2$, instead of the
naively expected and catastrophic $1/\aexp^2$, $\aexp$ being the scale
factor.  Nevertheless, overclosing of the universe is avoided only if
the energy density in the form of loops is radiated away. This is
typically the case for oscillating loops by means of gravitational
wave emission but may no longer be true if the strings carry particle
currents~\cite{Witten:1985eb, Davis:1988ip, Brandenberger:1996zp,
Ringeval:2001xd}.

On sub-horizon sizes, the understanding of string networks is still a
matter of debate. Early analytical studies predicted the scaling
property of long strings by considering the network dominated by one
length scale only: the inter-string distance, somehow related to the
horizon size~\cite{Kibble:1985hp,Albrecht:1989mk}. However, the first
high resolution numerical studies identified a copious production of
small sized loops, suggesting that the network dynamics should also
involve a small fundamental length scale~\cite{Bennett:1987vf,
Bennett:1989ak, Sakellariadou:1990nd, Bennett:1990yp,Allen:1990tv}.
This led to the development of the so-called three scales
model~\cite{Austin:1993rg}. In addition to two horizon sized
distances, the long strings develop, through their intersections, a
small scale wiggliness structure, explaining the formation of the
small sized loops. An interesting feature of the model is that the
small length scale reaches a scaling regime only if gravitational
radiation is considered, otherwise, and under reasonable assumptions,
the kinky structure is expected to incessantly grow with respect to
the horizon size. The main features of the three-scales model have
been numerically confirmed in Minkowski
space-time~\cite{Vincent:1996rb}. However, in these simulations,
nearly all loops are produced at the lattice spacing size, which makes
the evolution and the scaling properties of the small scale structure
strongly dependent on the cutoff. It was thus pointed out that if this
feature persists whatever the lattice spacing, the typical size of
physical loops might be the string width, which implies that particle
production rather than gravitational radiation would be the dominant
mode of energy dissipation from a string
network~\cite{Vincent:1996rb,Vincent:1997cx}. Although loop formation
at the string width has been recovered in more recent field numerical
simulations, it has been argued that loop production rather than
particle emission should remain the dominant cosmological energy
carrying mechanism~\cite{Avelino:1998qd,Moore:2001px}.

In the following, we present new results coming from numerical
simulations of cosmic strings evolution in a
Friedman--Lema\^{\i}tre--Robertson--Walker (FLRW) universe. For this
purpose, we used an improved version of the D.~Bennett and F.~Bouchet
Nambu-Goto cosmic string code~\cite{Bennett:1990yp}. We show for the
first time evidence of a scaling regime for the cosmic string loops in
both the radiation and matter eras down to a few thousandths of the
horizon size. The loops scaling evolution is similar to the long
strings one and does not rely on any gravitational back reaction
effect. It only appears after a relaxation period which is driven by a
transient overproduction of loops, with respect to the scaling value,
whose length is close to the initial correlation length of the string
network. We then discuss the effects induced by the finite numerical
resolution and show that they do not affect the loops scaling
regime. Moreover we confirm an explosive-like formation of very small
sized and numerically unresolved loops during the first stage of the
simulations, suggesting that particle production may briefly
dominate the physical evolution of a string network soon after its
formation.


\begin{figure}
\begin{center}
\includegraphics[width=14cm]{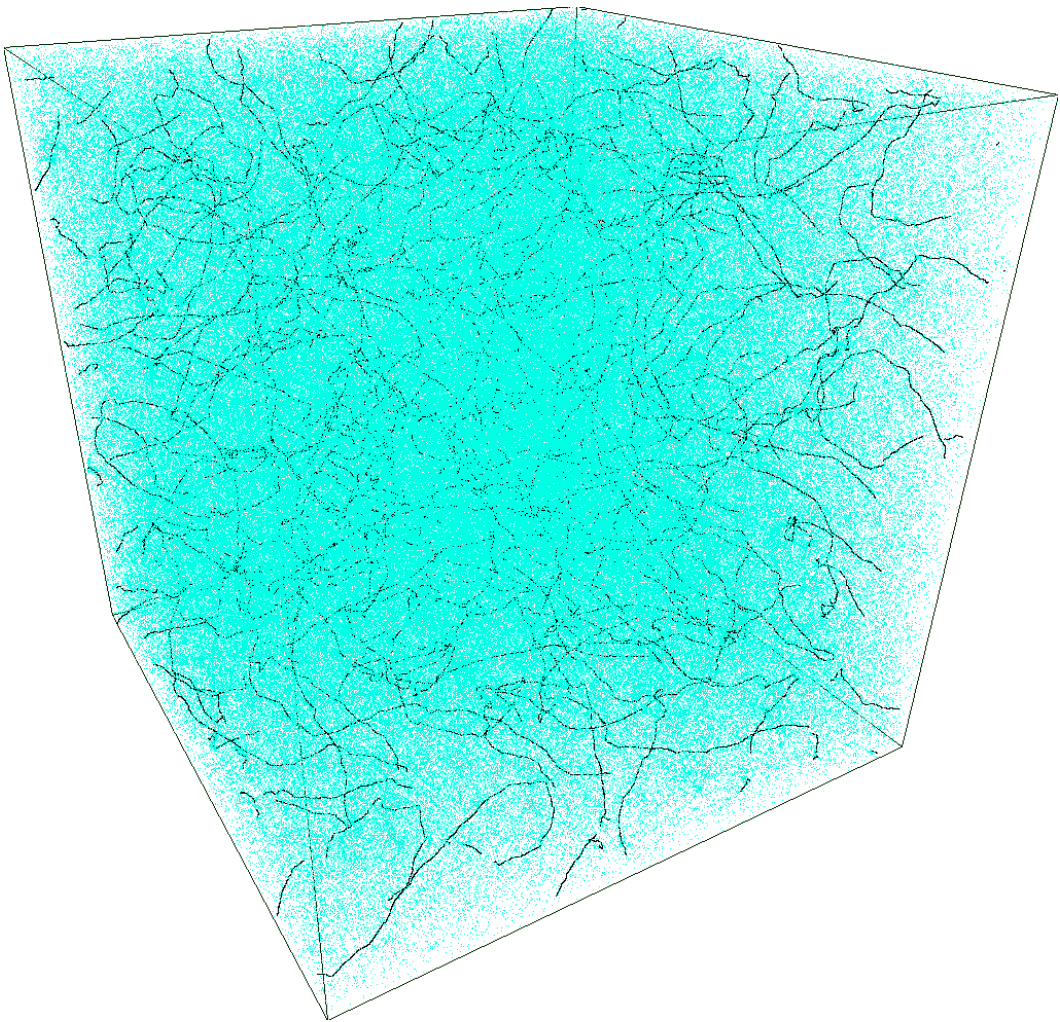}
\caption{The $(100\corrini)^3$ comoving volume in the matter era when
  the observable universe occupies one eighth of the box.}
\label{fig:snaprad}
\end{center}
\end{figure}

Our numerical simulations of strings in FLRW space-time are performed
in a fixed unity comoving volume with periodic boundary
conditions. The initial scale factor is normalised to unity while the
initial horizon size is a free parameter which controls the starting
string energy within a horizon volume. During the computations, the
comoving horizon size grows and the evolution is stopped before it
fills the whole unit volume for which the finiteness of the numerical
box starts to be felt. We used the Vachaspati--Vilenkin (VV) initial
conditions where the long strings path is essentially a random walk of
correlation length $\corrini$, together with a random transverse
velocity component of root mean squared amplitude
$0.1$~\cite{Vachaspati:1984dz,Bennett:1990yp}. Such a velocity
component allows a faster relaxation of the string network toward its
stable cosmological configuration. The results presented
below mostly come from two high resolution runs in the matter and
radiation eras, respectively, performed in a $(100\corrini)^3$
comoving box and with an initial string sampling of $20$ points per
correlation length (ppcl). The initial size of the horizon is chosen
to reduce the relaxation time toward the scaling regime and has been
set equal to $\horizonini=0.063$ for the matter era and
$\horizonini=0.057$ for the radiation era. The dynamic range achieved
by these runs is $8$ and $17$ in conformal time, which is equivalent
to $520$ and $308$ in physical time, for the matter and radiation era,
respectively (see Fig.~\ref{fig:snaprad}).

\begin{figure}
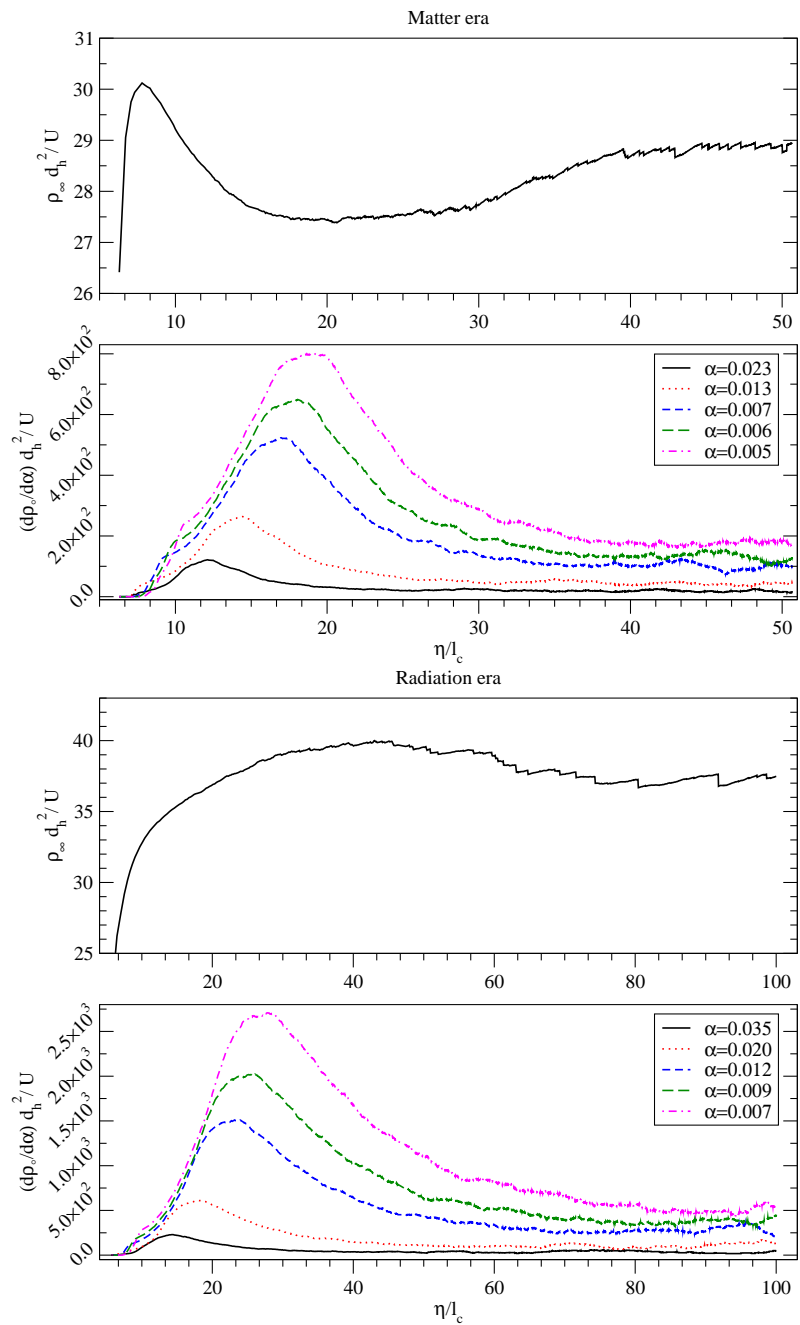

\begin{center}
\includegraphics[width=10.5cm]{enermat.eps}
\includegraphics[width=10.5cm]{enerrad.eps}
\caption{Evolution in the matter and radiation eras of the energy
  density associated with long strings and the energy density of loops
  of physical size $\lphys = \frachor \horizon$. The time variable is
  the rescaled conformal time $\eta/\corrini$. The energy
  density of long strings rapidly reaches a scaling regime where
  $\rho_\infty \propto 1/\horizon^2$, up to some damped relaxation
  oscillations. These plots show that the energy density of
  $\lphys$-sized loops also reaches a long string-like scaling regime
  where $\ud \enerloop \propto 1/\horizon^2$. This loop scaling regime
  appears after a relaxation period during which the energy density
  increases and decreases (the bumps). This transient energy excess
  signs the relaxation of the initial string network towards its
  cosmological stable configuration. Note that the transient regime is
  longer for the smaller loops, as expected from the hierarchical
  process of loop formation.}
\label{fig:enerscal}
\end{center}
\end{figure}

In Fig.~\ref{fig:enerscal}, we have plotted the evolution of the
energy density $\enerlong$ associated with the long strings,  defined
to be super-horizon sized, together with the loop energy density
distribution $\ud \enerloop/\ud \frachor$. More precisely, $\ud
\enerloop(\frachor)$ is the energy density carried by the loops whose
length, expressed in units of the horizon size, $\horizon$, is in the
range $\frachor$ to $\frachor + \ud \frachor$.  A logarithmic binning
in $\frachor$ of resolution $\Delta \frachor/\frachor \simeq 10^{-1}$
has been used in the range $\left[10^{-5},10^{2}\right]$ to compute
these quantities. From Fig.~\ref{fig:enerscal}, the scaling values
for the energy density associated with the infinite strings are
\begin{equation}
\label{eq:infscaling}
\enerlong \dfrac{\horizon^2}{\tension} \underset{\umat}{=} 28.4 \pm 0.9,
\qquad \enerlong \dfrac{\horizon^2}{\tension} \underset{\urad}{=}
37.8 \pm 1.7,
\end{equation}
for the matter and radiation era runs, where $\tension$ denotes the
string mass per unit length. The central value quoted above has been
defined to be the mean value in the last half conformal dynamic range
of each run while the errors correspond to the dispersions around the
mean during that time. These scaling values are compatible with
previous results while being slightly lower in the radiation era,
possibly due to our better numerical accuracy and dynamic range
~\cite{Bennett:1989ak,Allen:1990tv}. The really new results concern
the loop energy density distribution which, after a transient regime
(the bump in Fig.~\ref{fig:enerscal}), reaches a self-similar
evolution where $\ud \enerloop(\frachor) \horizon^2$ remains
stationary, for all values of $\frachor$ down to approximately $5
\times 10^{-3}$. The existence of such a loops scaling regime implies
that
\begin{equation}
\label{eq:scaling}
\dfrac{\ud \enerloop}{\ud \frachor} = \scaling(\frachor)
\dfrac{U}{\horizon^2}, \qquad \dfrac{\ud\numloop}{\ud
\frachor} = \dfrac{\scaling(\frachor)}{\frachor \, \horizon^3},
\end{equation}
where $\ud \numloop(\frachor)/\ud \frachor$ is the number density
distribution of loops. For a flat FLRW space-time, recall that
$\horizon=3t$ in the matter era and $\horizon=2t$ in the radiation
era. Notice that the loops smaller than $5 \times 10^{-3}$ the
size of the horizon have not yet entered such a scaling behaviour at
the end of the runs. As can be seen in Fig.~\ref{fig:enerscal}, the
transient regime characterised by the bump in the energy density
evolution is longer for the smaller loops. However, it is reasonable
to believe that these loops would have reached the scaling regime
with more dynamical range. These relaxation effects and their
relationship with the initial conditions are discussed later in the
letter.

\begin{figure}
\begin{center}
\includegraphics[width=15cm]{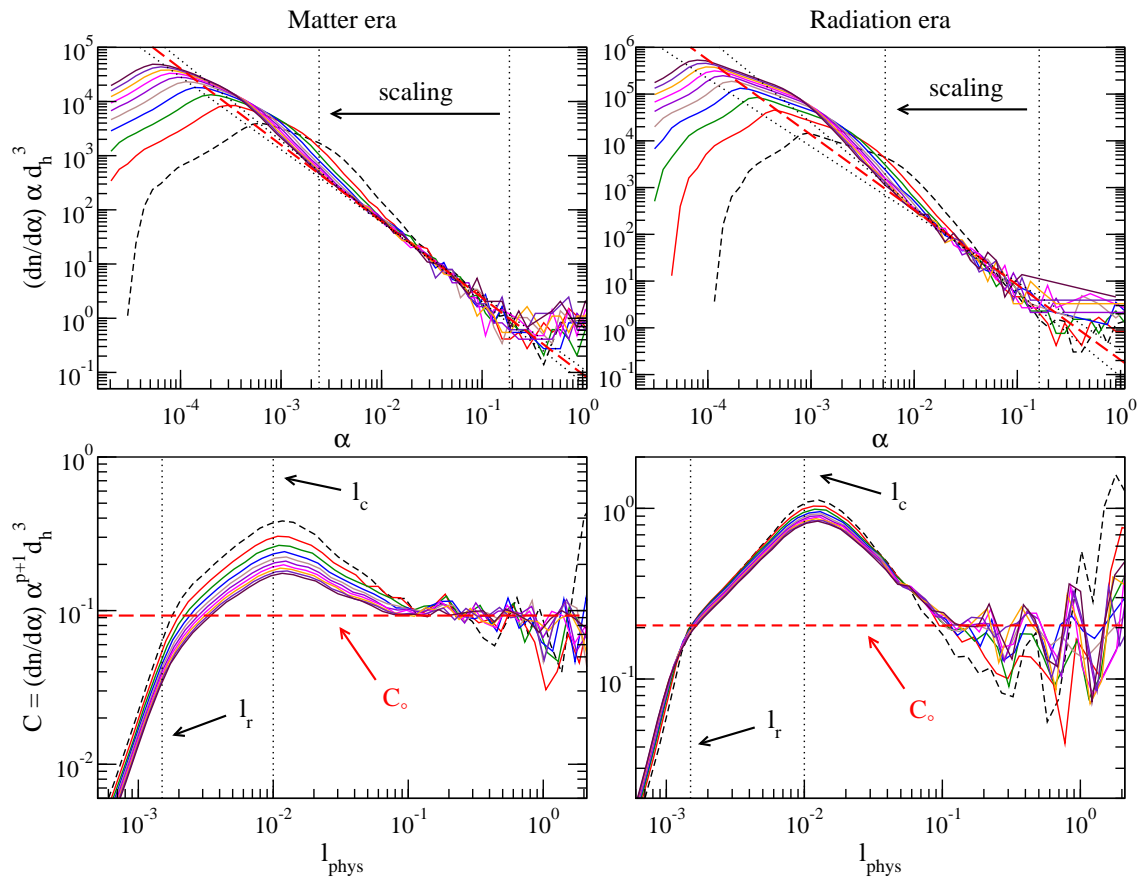}
\caption{The rescaled loop number density distributions with respect
  to $\frachor$ (top panels) and $\lphys=\frachor \horizon$ (bottom
  panels). They are plotted for different times starting at
  $\tphys=1.1$ and $\tphys=0.8$ (short dashed curves), with a physical
  time sampling equal to $1.1$ and $0.8$ for the matter and radiation
  runs, respectively. The scaling regime propagates from the large
  scales toward the small scales, while the relaxation bump around the
  initial correlation length $\corrini$ is progressively damped. The
  $\frachor$ intervals used to determine the scaling function in
  Eq.~(\ref{eq:powerlaw}) are represented, as well as the best power
  law fit (long dashed line) and its estimated systematic errors
  (dotted lines). Note that the overall maxima of the loop
  distributions in the top panels correspond to the knee centered
  around the initial resolution length $\resini$ in the bottom
  panels.}
\label{fig:numscal}
\end{center}
\end{figure}

In order to determine the scaling function $\scaling(\frachor)$ we
have plotted in the top part of Fig.~\ref{fig:numscal} the rescaled
distribution $ \frachor \horizon^3 \ud\numloop/\ud\frachor$ as a
function of $\frachor$, and at equally spaced physical times spread
over the dynamic range of the simulations.  As expected from the
establishment of scaling observed in Fig.~\ref{fig:enerscal}, the
distribution functions in Fig.~\ref{fig:numscal} start to superimpose
at the largest length scales during earlier times. Simultaneously, the
non-scaling parts of the distribution function shift toward small
$\frachor$. This essentially means that the intersection processes
which create loops from long strings take place in a very similar
manner for loops themselves.  Their self-intersections give rise to
more numerous smaller loops in such a way that a constant energy flow
cascades from the long strings to the smallest loops. Conversely, due
to their small size, the reconnection processes of these freshly formed
smaller loops to long strings and large loops is
negligible~\cite{Bennett:1990yp}. As a result, since the source of
energy is the entering of long strings inside the horizon size, it is
not really surprising that the horizon ends up being the relevant
quantity also for the loops. Note that some hints of these effects
have recently been observed on the kinky structure of strings in
Minkowski space-time~\cite{Vanchurin:2005yb}.

 We have performed a power law least squares fit of $\frachor
\horizon^3 \ud\numloop/\ud\frachor$ in the domain $\frachorscal <
\frachor <\frachorlong$ where the loops scaling regime has been
reached. Our prescription is to set $\frachorscal$ as the lowest size
of loops, in units of the horizon size, for which the corresponding
energy density $\ud \enerloop(\frachorscal)$ remains stationary during
the last $5\%$ of the simulation conformal time range (see
Fig.~\ref{fig:enerscal} and Fig.~\ref{fig:numscal}). In order to avoid
contamination by long string effects, the upper bound has been fixed
to the typical distance between infinite strings, \ie
$\frachorlong=(\tension/\enerlong)^{1/2} / \horizon$. By fitting the
data points averaged over the above-mentioned $5\%$ conformal time
interval, one gets the power law fit $\scaling(\frachor) = \const \,
\frachor^{-\power}$ with
\begin{equation}
\label{eq:powerlaw}
\begin{array}{cc}
\left\{
\begin{array}{ccc}
\power & \underset{\umat}{=} & 1.41\,^{+0.08}_{-0.07} \\
\const & \underset{\umat}{=} & 0.09\,^{-0.03}_{+0.03}
\end{array}
\right.,
\quad \mathrm{and} \quad
\left\{
\begin{array}{ccc}
\power & \underset{\urad}{=} & 1.60\,^{+0.21}_{-0.15} \\
\const & \underset{\urad}{=} & 0.21\,^{-0.12}_{+0.13}
\end{array}
\right.,
\end{array}
\end{equation}
for the matter and radiation run, respectively. The errors quoted
above mainly come from systematic errors associated with the
determination of the $\frachor$ domains where the loops are in the
scaling regime. They have been estimated as the largest $\const$ and
$\power$ deviations between four power law fits. These other
fits use the same data but involve the four $\frachor$ intervals which
are obtained by multiplying and dividing each of the original bounds
$\frachorscal$ and $\frachorlong$ by a factor of two.

\begin{figure}
\begin{center}
\includegraphics[width=15cm]{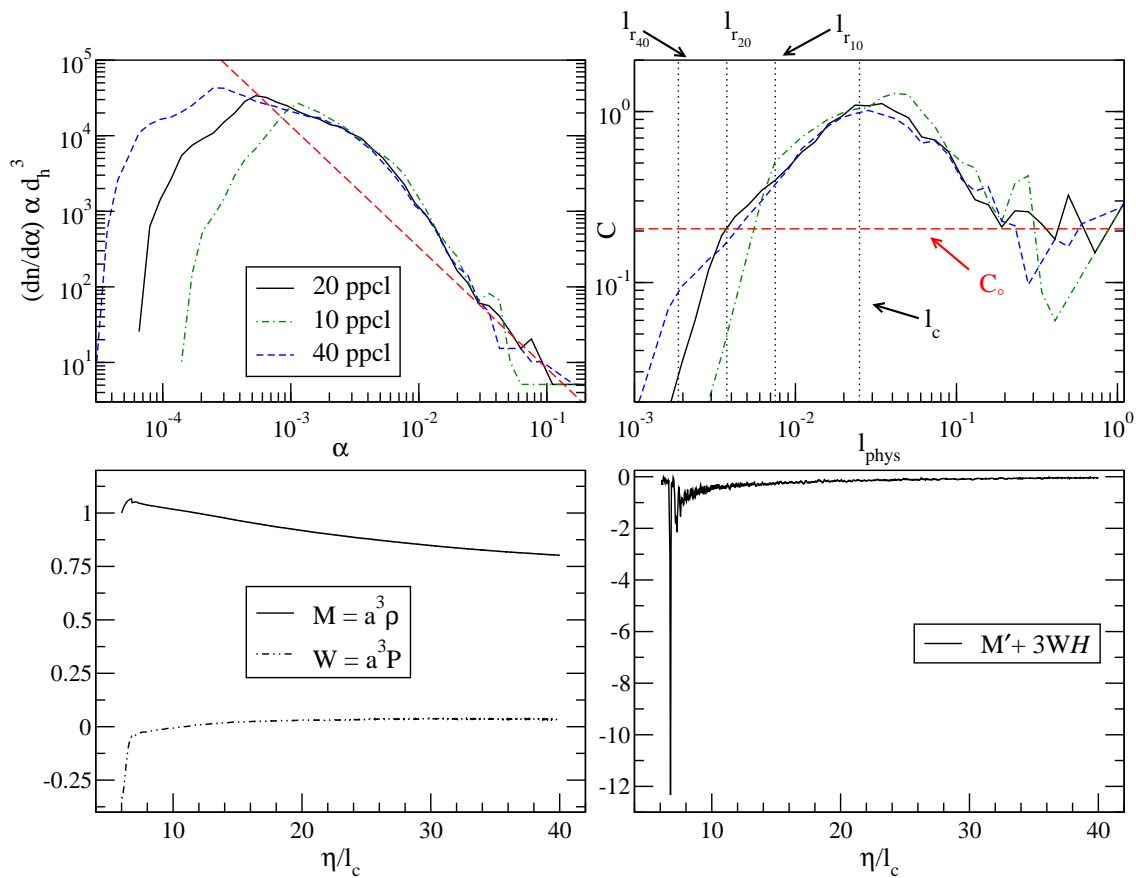}
\caption{Influence of the initial resolution length $\resini$ on the
  rescaled loop distributions at the end of three small $(40
  \corrini)^3$ radiation era runs having an initial sampling of $10$,
  $20$ and $40$ ppcl, and a dynamic range of $45$ in physical time
  (top panels). As in the previous figures, $\frachor =
  \lphys/\horizon$ is the loop length in units of the horizon size. In
  the bottom panels, the influence of the $3$ points cutoff is
  quantified for the $20$ ppcl run by checking for stress energy
  conservation.}
\label{fig:dissip}
\end{center}
\end{figure}

As can be seen in the top panels of Fig.~\ref{fig:numscal}, the
transient overproduction of loops preceding the scaling (the bump
right to maximum) and the overall maximum of the loop distribution
evolve in time. During the runs, they peak at decreasing sizes with
respect to the horizon size. In the bottom panels of
Fig.~\ref{fig:numscal}, we have plotted the rescaled distributions
$\noconst \equiv \frachor^{p+1} \horizon^3 \ud \numloop/\ud \frachor$
for the matter and radiation eras. For the length scales in scaling,
these distributions are approximated by the above-mentioned constant
$\const$ whereas the relaxation regime and other non-scaling features
appear as deviations from this flat distribution. Note that, this
time, we have plotted the rescaled distributions with respect to the
physical length $\lphys=\frachor\horizon$. They now peak around a
constant value which is close to the initial physical correlation
length $\corrini=1/100$ associated with the VV initial conditions. The
setting up of the scaling regime appears in this frame as the
progressive damping of the bump, which flattens the distributions
toward $\const$ and smooths out the correlations coming from the
initial conditions. Moreover, the overall maximum of the $\frachor
\horizon^3 \ud \numloop/\ud \frachor$ distribution appears as a knee,
which also remains at a constant physical length during the
simulation. This length ends up being close to the initial resolution
length associated with the VV initial conditions (see also
Fig.~\ref{fig:dissip}). As detailed in Ref.~\cite{Bennett:1990yp}, our
numerical simulations are not lattice size limited in physical space
due to an adaptive griding on each loop. However, at any time, loops
with strictly less than $3$ points cannot be formed and this is
especially relevant for the initial conditions. Indeed, for an initial
sampling of $20$ ppcl, this cutoff necessarily leads to additional
correlations in the VV network around the physical length
$\resini=(3/20)\corrini$, which precisely corresponds to the knee of
the rescaled loop distributions $\noconst$. The fact that the
correlations associated with both $\corrini$ and $\resini$ remain at
constant physical lengths during the subsequent evolution may be
compared to the evolution of the small-scale correlation length
obtained in the three-scale models in absence of gravitational damping
and under other reasonable assumptions~\cite{Austin:1993rg}.


Since the above-mentioned discretisation effects concern the smallest
loops, they should not influence the properties of the string network
on larger scales. In the top panels of Fig.~\ref{fig:dissip}, we have
plotted the rescaled loop distributions obtained at the end of three
$(40 \corrini)^3$ radiation era runs having an initial sampling of
$10$, $20$ and $40$ ppcl, respectively. As expected, one can check
that the finite resolution effects remain confined to length scales
smaller than the initial correlation length $\corrini$ of the string
network and do not affect the previously discussed loop scaling
regime. We have also checked for these smaller runs the
insensitivity of the loop distribution with respect to the initial
random velocity component.

In previous studies of Nambu-Goto strings, the dissipation effects
have been associated with the formation of lattice sized loops in
physical space~\cite{Vincent:1996rb,Vincent:1997cx}. In the context of
our $3$ points cutoff, all triangle shaped loops are removed from the
subsequent evolution since they cannot self-intersect and do not
significantly alter the evolution of larger scales, but would
otherwise require additional computing time. Since such a removal is
not equivalent to a fixed physical size cutoff, another way to discuss
the dissipation effects is to test the total stress energy
conservation during the evolution. Designating by $\enertot$ and
$\prestot$ the total string energy density and pressure, we have
plotted in the bottom panels of Fig.~\ref{fig:dissip} the evolution of
the total network mass $\masstot = a^3 \enertot$, the total pressure
work $\worktot=a^3 \prestot$, and the energy dissipation rate
$\losttot \equiv \masstot' + 3 \hubbleconf \worktot$ for the
$(40\corrini)^3$ radiation era run ($20$ ppcl). The prime stands for
the derivative with respect to the conformal time and
$\hubbleconf=a'/a$ is the conformal Hubble parameter. From the
conservation of the Nambu-Goto stress tensor, one would expect
$\losttot=0$. As a result, the existence of a sharp negative peak for
$\losttot$ at the very beginning of the run signals a strong energy
loss rate in the form of numerically unresolved loops (see
Fig.~\ref{fig:dissip}). However, this happens only during a brief
period in which the universe expands less than a factor of $10^{-3}$
and corresponds to the times at which the overall maximum of the loop
distribution around the initial resolution length appears. As the
pressure evolution shows, this observation confirms the existence of
an explosive-like relaxation of the VV strings network toward the
cosmological network. Since in the early universe a string network
soon after its formation is certainly far from its cosmological stable
configuration, one might also expect such an explosive relaxation to
physically take place. Note that during the later times of the
numerical simulation the stress energy conservation is quite well
satisfied, especially during the loop scaling regime.


In summary, we have numerically shown that the energy and number
densities of cosmic string loops in an expanding universe reach a
scaling regime down to a few thousandths of the horizon size. This
result does not rely on gravitational back reaction effects and
suggests that the loop formation mechanism is the dominant energy
carrying mechanism in the cosmological context. The loops scaling
regime takes place after a relaxation period during which we observe
an overproduction of loops peaked around a constant physical length
close to the initial correlation length $\corrini$ of the string
network.  We have confirmed the existence of a brief explosive-like
dissipative behaviour in the very first stage of the evolution where a
few percent of the total energy density are lost in the form of
numerically unresolved loops. At the same time, the overall maximum of
the loop distribution appears around a constant physical length scale
close to the initial resolution length $\resini$ of the string
network. These effects suggest that particle production may dominate
the relaxation of a string network in the early universe from the
string forming phase transition toward its cosmological scaling
configuration. We have however shown that these discretisation effects
do not affect the loops scaling regime.

Our results should provide an improved basis for the development of
loop evolution models~\cite{Copeland:1998na} and for the determination
of the gravitational wave background associated with the existence of
a cosmological string network~\cite{Damour:2004kw}. On the other
hand, our Nambu-Goto string model certainly breaks down for length
scales close to string width. One therefore expect the cosmological
loop distribution given in Eq.~(\ref{eq:scaling}) to be valid above a
given cutoff fixed by physical processes not included in our analysis.

\acknowledgments

We thank Tom Kibble, Carlos Martins, Patrick Peter, Danielle Steer and
Pascal Voury for various enlightening and motivating discussions.
Special thanks to D.~Bennett who permitted us to resuscitate and use
the code as the basis of this article. The computations have been
performed on an IBM Power4 P690+ machine during seven CPU-years at the
``Institut du D\'eveloppement des Ressources en Informatique
Scientifique'' (\texttt{http://www.idris.fr}). The work of M.~S. was
supported in part by the European Union through the Marie Curie
Research and Training Network UniverseNet (MRTN-CN-2006-035863).

\section*{Note added}

A week after our results were rendered public on the archives, two
other papers dealing with the scaling of cosmic string loops appeared:
Refs~\cite{Vanchurin:2005pa} and \cite{Martins:2005es}. Each of these
works exhibits its own loop distribution and we would like to briefly
discuss qualitatively what we suspect to be at the origin of these
differences.

In Ref.~\cite{Vanchurin:2005pa}, the authors studied cosmic string
evolution under the simplifying hypothesis of Minkowski spacetime and
observed that most of their primary loops are produced at a size
$0.3t$ before going into loops of scaling size $0.1t$, a value close
to their long inter-string distance. In order to attempt a direct
comparison between Minkowski and FLRW simulations, one may express
lengths with respect to the typical inter-string distance (our
$\alpha_\infty$): their primary loop production size $0.3t$ would
correspond to our $\alpha\simeq 0.5$ (for the matter era). As can be
seen in our Fig.~\ref{fig:numscal}, this is out of our scaling
range. Nevertheless, making abstraction of the statistical
fluctuations induced by the small number of loops we have on that
scale, a peak might be guessed, at least for the matter era run. As
far as we can compare these results, they seem to be associated with
loop formation events typical of the one-scale Kibble model. On the
other hand, their loop distribution at small scales differs from ours
and we notably do not observe the equivalent of their peak around
$0.1t$. In our opinion, this may be related to their numerical cutoff
which switches off the loop dynamics in the network as soon as their
size is smaller than $0.25t$. We do not use any similar
assumption\footnote{More recently, Ref.~\cite{Olum:2006ix} presented
some new results by using the FLRW integration scheme of
Ref.~\cite{Bennett:1990yp}. Their loop distributions are still
different from ours albeit closer. The $0.25t$ cutoff has been
replaced by a removal of all loops which do not intersect after one
oscillation.}. Let us also notice that their long dynamical range is
obtained at the expense of a procedure to increase the simulation
volume, a procedure which may introduce spurious correlations on all
length scales and which is hardly tractable in FLRW spacetime.

In Ref.~\cite{Martins:2005es}, the authors performed the numerical
simulations in FLRW spacetime and their results are more directly
comparable to ours. They used a high number of points per correlation
length (ppcl) to sample the strings and therefore have a smaller
dynamical range than us. As a result, the loops we have found to be in
scaling might still be in the transient regime at the end of their
runs (see our Fig.~\ref{fig:enerscal}). The number of ppcl has two
effects. Firstly, high resolution ensures high numerical accuracy to
describe the Nambu-Goto dynamics, \ie the numerically computed
worldsheet evolution remains closer to the true solution for a given
algorithm. As thoroughly discussed in Ref.~\cite{Bennett:1990yp}
(Sect.~IIB-3), the integration method we used reaches the same
numerical accuracy as the other codes but with a factor of three to
four less points per correlation length. The second effect of large
ppcl is to reduce the initial resolution length $\resini$, and
consequently the typical size of the purely numerical
extra-correlations added in the initial string network by the
discretisation. There is no mention in Ref.~\cite{Martins:2005es} of
such kind of effect. However, their Fig.~4 is almost identical to our
loop distribution of Fig.~\ref{fig:numscal} but plotted against
$\ell/t$. From this figure, they claim that the peak location of their
loop distribution is moving slower at late time as if the whole
distribution were approaching a scaling regime. In our interpretation,
since we have found that the overall peak is precisely located at
$\resini$, a fixed physical length, such a slowing down solely comes
from the choice of variable $\ell/t$ (see our Fig.~\ref{fig:numscal},
bottom panels).

In spite of these differences, both of these works claimed to observe
some scaling evolution for the cosmic string loops, independently of
radiative back reaction effects, thereby confirming our main result.

\section*{References}

\bibliography{bibstring}

\providecommand{\href}[2]{#2}\begingroup\raggedright\begin{thebibliography}{10}

\bibitem{kibble:1976}
T.~Kibble, {\it Topology of cosmic domains and strings},  {\em J. Phys.} {\bf
  A9} (1976) 1387.

\bibitem{kibble:1980}
T.~Kibble, {\it Some implications of a cosmological phase transition},  {\em
  Phys. Rept.} {\bf 67} (1980) 183.

\bibitem{Jeannerot:2003qv}
R.~Jeannerot, J.~Rocher, and M.~Sakellariadou, {\it How generic is cosmic
  string formation in susy guts},  {\em Phys. Rev.} {\bf D68} (2003) 103514,
  [\href{http://xxx.lanl.gov/abs/hep-ph/0308134}{{\tt hep-ph/0308134}}].

\bibitem{Vilenkin:2000}
A.~Vilenkin and E.~Shellard, {\em Cosmic Strings and Other Topological
  Defects}.
\newblock Cambridge University Press, Cambridge, UK, 2000.

\bibitem{Hindmarsh:1994re}
M.~B. Hindmarsh and T.~W.~B. Kibble, {\it Cosmic strings},  {\em Rept. Prog.
  Phys.} {\bf 58} (1995) 477--562,
  [\href{http://xxx.lanl.gov/abs/hep-ph/9411342}{{\tt hep-ph/9411342}}].

\bibitem{Sarangi:2002yt}
S.~Sarangi and S.~H.~H. Tye, {\it Cosmic string production towards the end of
  brane inflation},  {\em Phys. Lett.} {\bf B536} (2002) 185--192,
  [\href{http://xxx.lanl.gov/abs/hep-th/0204074}{{\tt hep-th/0204074}}].

\bibitem{Copeland:2003bj}
E.~J. Copeland, R.~C. Myers, and J.~Polchinski, {\it Cosmic f- and d-strings},
  {\em JHEP} {\bf 06} (2004) 013,
  [\href{http://xxx.lanl.gov/abs/hep-th/0312067}{{\tt hep-th/0312067}}].

\bibitem{Dvali:2003zj}
G.~Dvali and A.~Vilenkin, {\it Formation and evolution of cosmic d-strings},
  {\em JCAP} {\bf 0403} (2004) 010,
  [\href{http://xxx.lanl.gov/abs/hep-th/0312007}{{\tt hep-th/0312007}}].

\bibitem{Witten:1985eb}
E.~Witten, {\it Superconducting strings},  {\em Nucl. Phys.} {\bf B249} (1985)
  557--592.

\bibitem{Davis:1988ip}
R.~L. Davis, {\it Semitopological solitons},  {\em Phys. Rev.} {\bf D38} (1988)
  3722.

\bibitem{Brandenberger:1996zp}
R.~H. Brandenberger, B.~Carter, A.-C. Davis, and M.~Trodden, {\it Cosmic
  vortons and particle physics constraints},  {\em Phys. Rev.} {\bf D54} (1996)
  6059--6071, [\href{http://xxx.lanl.gov/abs/hep-ph/9605382}{{\tt
  hep-ph/9605382}}].

\bibitem{Ringeval:2001xd}
C.~Ringeval, {\it Fermionic massive modes along cosmic strings},  {\em Phys.
  Rev.} {\bf D64} (2001) 123505,
  [\href{http://xxx.lanl.gov/abs/hep-ph/0106179}{{\tt hep-ph/0106179}}].

\bibitem{Kibble:1985hp}
T.~W.~B. Kibble, {\it Evolution of a system of cosmic strings},  {\em Nucl.
  Phys.} {\bf B252} (1985) 227.

\bibitem{Albrecht:1989mk}
A.~Albrecht and N.~Turok, {\it Evolution of cosmic string networks},  {\em
  Phys. Rev.} {\bf D40} (1989) 973--1001.

\bibitem{Bennett:1987vf}
D.~P. Bennett and F.~R. Bouchet, {\it Evidence for a scaling solution in cosmic
  string evolution},  {\em Phys. Rev. Lett.} {\bf 60} (1988) 257.

\bibitem{Bennett:1989ak}
D.~P. Bennett and F.~R. Bouchet, {\it Cosmic string evolution},  {\em Phys.
  Rev. Lett.} {\bf 63} (1989) 2776.

\bibitem{Sakellariadou:1990nd}
M.~Sakellariadou and A.~Vilenkin, {\it Cosmic-string evolution in flat
  space-time},  {\em Phys. Rev.} {\bf D42} (1990) 349--353.

\bibitem{Bennett:1990yp}
D.~P. Bennett and F.~R. Bouchet, {\it High resolution simulations of cosmic
  string evolution. 1. network evolution},  {\em Phys. Rev.} {\bf D41} (1990)
  2408.

\bibitem{Allen:1990tv}
B.~Allen and E.~P.~S. Shellard, {\it Cosmic string evolution: A numerical
  simulation},  {\em Phys. Rev. Lett.} {\bf 64} (1990) 119--122.

\bibitem{Austin:1993rg}
D.~Austin, E.~J. Copeland, and T.~W.~B. Kibble, {\it Evolution of cosmic string
  configurations},  {\em Phys. Rev.} {\bf D48} (1993) 5594--5627,
  [\href{http://xxx.lanl.gov/abs/hep-ph/9307325}{{\tt hep-ph/9307325}}].

\bibitem{Vincent:1996rb}
G.~R. Vincent, M.~Hindmarsh, and M.~Sakellariadou, {\it Scaling and small scale
  structure in cosmic string networks},  {\em Phys. Rev.} {\bf D56} (1997)
  637--646, [\href{http://xxx.lanl.gov/abs/astro-ph/9612135}{{\tt
  astro-ph/9612135}}].

\bibitem{Vincent:1997cx}
G.~Vincent, N.~D. Antunes, and M.~Hindmarsh, {\it Numerical simulations of
  string networks in the abelian- higgs model},  {\em Phys. Rev. Lett.} {\bf
  80} (1998) 2277--2280, [\href{http://xxx.lanl.gov/abs/hep-ph/9708427}{{\tt
  hep-ph/9708427}}].

\bibitem{Avelino:1998qd}
P.~P. Avelino, E.~P.~S. Shellard, J.~H.~P. Wu, and B.~Allen, {\it Cosmic string
  loops and large-scale structure},  {\em Phys. Rev.} {\bf D60} (1999) 023511,
  [\href{http://xxx.lanl.gov/abs/astro-ph/9810439}{{\tt astro-ph/9810439}}].

\bibitem{Moore:2001px}
J.~N. Moore, E.~P.~S. Shellard, and C.~J. A.~P. Martins, {\it On the evolution
  of abelian-higgs string networks},  {\em Phys. Rev.} {\bf D65} (2002) 023503,
  [\href{http://xxx.lanl.gov/abs/hep-ph/0107171}{{\tt hep-ph/0107171}}].

\bibitem{Vachaspati:1984dz}
T.~Vachaspati and A.~Vilenkin, {\it Formation and evolution of cosmic strings},
   {\em Phys. Rev.} {\bf D30} (1984) 2036.

\bibitem{Vanchurin:2005yb}
V.~Vanchurin, K.~Olum, and A.~Vilenkin, {\it Cosmic string scaling in flat
  space},  {\em Phys. Rev.} {\bf D72} (2005) 063514,
  [\href{http://xxx.lanl.gov/abs/gr-qc/0501040}{{\tt gr-qc/0501040}}].

\bibitem{Copeland:1998na}
E.~J. Copeland, T.~W.~B. Kibble, and D.~A. Steer, {\it The evolution of a
  network of cosmic string loops},  {\em Phys. Rev.} {\bf D58} (1998) 043508,
  [\href{http://xxx.lanl.gov/abs/hep-ph/9803414}{{\tt hep-ph/9803414}}].

\bibitem{Damour:2004kw}
T.~Damour and A.~Vilenkin, {\it Gravitational radiation from cosmic
  (super)strings: Bursts, stochastic background, and observational windows},
  {\em Phys. Rev.} {\bf D71} (2005) 063510,
  [\href{http://xxx.lanl.gov/abs/hep-th/0410222}{{\tt hep-th/0410222}}].

\bibitem{Vanchurin:2005pa}
V.~Vanchurin, K.~D. Olum, and A.~Vilenkin, {\it Scaling of cosmic string
  loops},  {\em Phys. Rev.} {\bf D74} (2006) 063527,
  [\href{http://xxx.lanl.gov/abs/gr-qc/0511159}{{\tt gr-qc/0511159}}].

\bibitem{Martins:2005es}
C.~J. A.~P. Martins and E.~P.~S. Shellard, {\it Fractal properties and
  small-scale structure of cosmic string networks},  {\em Phys. Rev.} {\bf D73}
  (2006) 043515, [\href{http://xxx.lanl.gov/abs/astro-ph/0511792}{{\tt
  astro-ph/0511792}}].

\bibitem{Olum:2006ix}
K.~D. Olum and V.~Vanchurin, {\it Cosmic string loops in the expanding
  universe},  \href{http://xxx.lanl.gov/abs/astro-ph/0610419}{{\tt
  astro-ph/0610419}}.

\end{thebibliography}\endgroup

\end{document}